\documentclass{aa}  

\usepackage{graphicx}
\usepackage{txfonts}
 \usepackage{todonotes}
\usepackage[T1]{fontenc} 
\usepackage{gensymb}
\usepackage[utf8]{inputenc}
\usepackage[T1]{fontenc}
\usepackage{ae,aecompl,graphicx,amsmath,xspace}
\usepackage{esvect}
\usepackage{mathrsfs}
\usepackage{xcolor,soul,afterpage,subcaption, xparse, xspace, csquotes}
\usepackage[skip=9pt]{caption}
\xspaceaddexceptions{]\}}
\allowdisplaybreaks
\usepackage{tikz}
\usetikzlibrary{shapes}
\usepackage{float}
\usepackage{hyperref}
\usepackage{appendix, apptools}
\newcommand{\refeq}[1]{Equation~(\ref{eq:#1})}   
       
\newcommand{\reffig}[1]{Figure~\ref{fig:#1}}
          
\newcommand{\refapp}[1]{Appendix~\ref{app:#1}}

\def\be{\begin{equation}}
\def\ee{\end{equation}}
\def\bea{\begin{eqnarray}}
\def\eea{\end{eqnarray}}

\def\ba#1\ea{\begin{align}#1\end{align}}
\def\bab#1\eab{\begin{equation}\begin{aligned}[b]#1\end{aligned}\end{equation}}

\def\bg#1\eg{\begin{gather}#1\end{gather}}

\usepackage{xcolor}
\hypersetup{
    colorlinks,
    linkcolor={red!50!black},
    citecolor={blue!50!black},
    urlcolor={blue!80!black}
}

\newcommand\Mpc{$\;\mathrm{Mpc}$\xspace}

\def\M{\mathcal{M}}
\def\N{\mathcal{N}}

\def\P{\mathcal{P}}

\newcommand\given[1][]{\:#1\vert\:}

\def\<{\left\langle}
\def\>{\right\rangle}

\def\P{\mathcal{P}}
\def\M{M^\mathrm{g}}
\def\N{N^\mathrm{g}}
\def\codefont#1{\texttt{#1}}
\def\emph#1{\textit{#1}}

\definecolor{nicegreen}{HTML}{2CA02C}
\definecolor{darkgreen}{rgb}{0.0,0.5,0.0}

\begin{document} 

   \title{Higher-order statistics of the large-scale structure from photometric redshifts}
   
   \titlerunning{Large-scale structure from photometric redshifts}
   
   \authorrunning{Tsaprazi et al.}

   \author{Eleni Tsaprazi$^1$, Jens Jasche$^{1,2}$, Guilhem Lavaux$^2$ and Florent Leclercq$^2$}

   \institute{The Oskar Klein Centre, Department of Physics, Stockholm University, Albanova University Center, SE 106 91 Stockholm, Sweden\\
              \email{eleni.tsaprazi@fysik.su.se}
         \and
             CNRS \& Sorbonne Université, UMR 7095, Institut d’Astrophysique de Paris, 98 bis boulevard Arago, F-75014 Paris, France\\
             }

   \date{}

 
  \abstract
   {The large-scale structure is a major source of cosmological information. However, next-generation photometric galaxy surveys will only provide a distorted view of cosmic structures due to large redshift uncertainties.}
   {To address the need for accurate reconstructions of the large-scale structure in presence of photometric uncertainties, we present a framework that constrains the three-dimensional dark matter density jointly with galaxy photometric redshift probability density functions (PDFs), exploiting information from galaxy clustering.}
   {Our forward model provides Markov Chain Monte Carlo realizations of the primordial and present-day dark matter density, inferred jointly from data. Our method goes beyond 2-point statistics via field-level inference. It accounts for all observational uncertainties and the survey geometry.}
   {We showcase our method using mock catalogs that emulate next-generation surveys with a worst-case redshift uncertainty, equivalent to ${\sim}300$\Mpc{}. On scales $150$\Mpc{}, we improve the cross-correlation of the photometric galaxy positions with the ground truth from $28\%$ to $86\%$. The improvement is significant down to $13$\Mpc{}. On scales $150$\Mpc{}, we achieve a cross-correlation of $80-90\%$ with the ground truth for the dark matter density, radial peculiar velocities, tidal shear and gravitational potential.}
   {We achieve accurate inferences of the large-scale structure on scales smaller than the original redshift uncertainty. Despite the large redshift uncertainty, we recover individual cosmic structures. Owing to our structure growth model, we infer plausible initial conditions of structure formation. Finally, we constrain individual photometric redshift PDFs. This work opens up the possibility to extract information at the smallest cosmological scales with next-generation photometric surveys, going beyond approaches that compress information in the data.}


   \maketitle
%
\section{Introduction}
\label{sec:intro}


High-accuracy galaxy redshift estimates are required for a plethora of scientific cases, such as the mapping of the cosmic large-scale structure and tests on the geometry of the universe \citep[e.g.][]{2006ApJ...636...21M,2006astro.ph..9591A,2006ewg3.rept.....P,10.1093/mnras/stu754,2017AAS...22922602M,2017MNRAS.465L..20S,2018arXiv180901669T,2018ARA&A..56..393M,2019ApJ...884..164Y,2019MNRAS.486.2730A,2021MNRAS.507.1746E,2021arXiv211006947L,2022PhRvD.105b3515S,2022MNRAS.511.1029H,2022arXiv220613633N}. Next-generation surveys will deliver low-accuracy redshift estimates from photometry and in certain cases fewer, high-accuracy, spectroscopic ones \citep[e.g.][]{2009arXiv0912.0201L,Euclid:2013,2014arXiv1412.4872D,HSC:2018,LSST:2019}.
However, bias in cosmological inferences may occur in sky areas, depth and color ranges where only photometric data is available \citep[e.g.][]{2006ApJ...636...21M}. Fortunately, the physical clustering information of the large-scale structure can be used to improve the accuracy of photometric redshift observations \citep[e.g.][]{1979ApJ...227...30S,1987MNRAS.229..621P,1996ApJ...460...94L,2010ApJ...708..505K,2012MNRAS.425.1042J,2013arXiv1303.4722M,2015MNRAS.454..463A,2020A&A...636A..90S,2021PhRvD.103d3520M}.

\begin{figure*}[h]
    \centering
    \includegraphics[width=\textwidth]{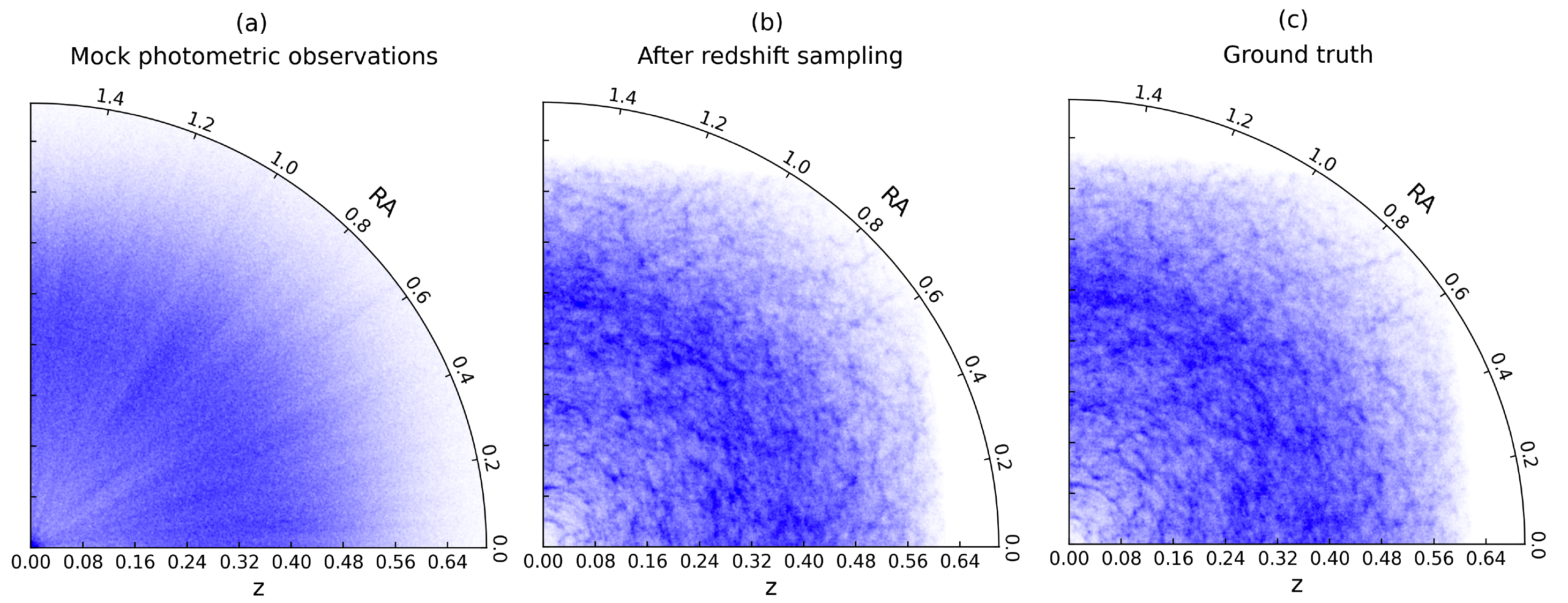}
    \caption{(a) Mock photometric galaxy positions. Galaxy positions are radially distorted due to redshift uncertainty. (b) Galaxy positions in a typical MCMC sample after the application of our method. Galaxies now trace the filamentary dark matter distribution. (c) Ground truth (mock) galaxy positions. The galaxy positions that were radially smeared in the mock observations, closely trace the filamentary structure in the ground truth galaxy positions -- within observational uncertainties -- after the application of our algorithm.}
    \label{fig:plot_2}
\end{figure*}

\begin{figure}[h]
    \centering
    \includegraphics[width=9cm]{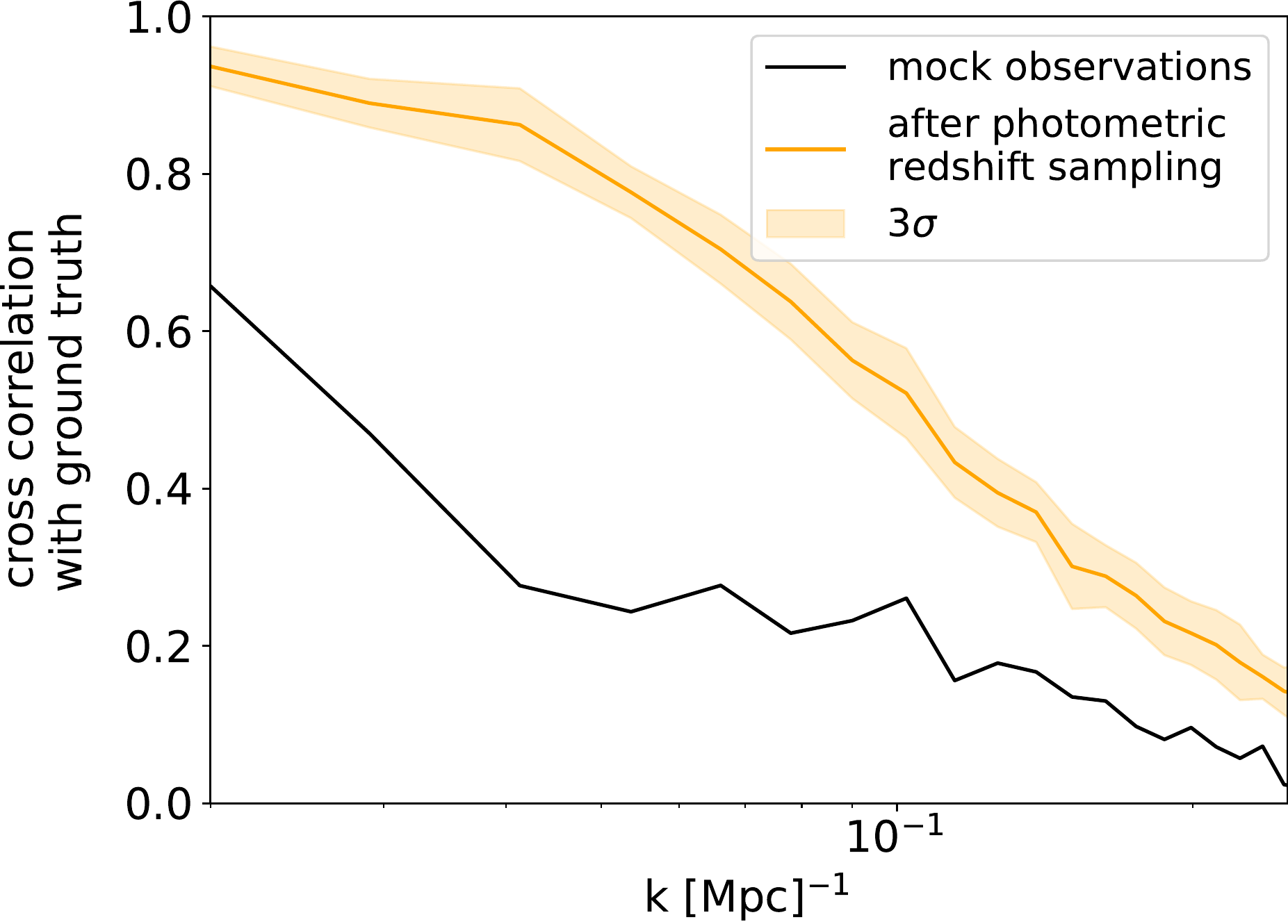}
    \caption{Cross-correlation of the gridded photometric galaxy coordinates before (black) and after (orange) the application of our method with the ground truth (mock data). The 3$\sigma$ error bars are across MCMC realizations. We consider a subbox that is the largest unmasked cubic volume in our inference domain, with side length $520$\Mpc{}. On the largest scales, $\sim 1.7\langle \sigma_z \rangle$, the cross-correlation increases from $73\%$ to $96\%$. On the smallest scales, $\sim 0.04\langle \sigma_z \rangle$, the cross-correlation increases from $2\%$ to $14\%$. The largest improvement is on scales $\sim 0.5\langle \sigma_z \rangle$, from $28\%$ to $86\%$.}
    \label{fig:plot_3}
\end{figure}

\begin{figure*}[h]
    \centering
    \includegraphics[width=0.9\textwidth]{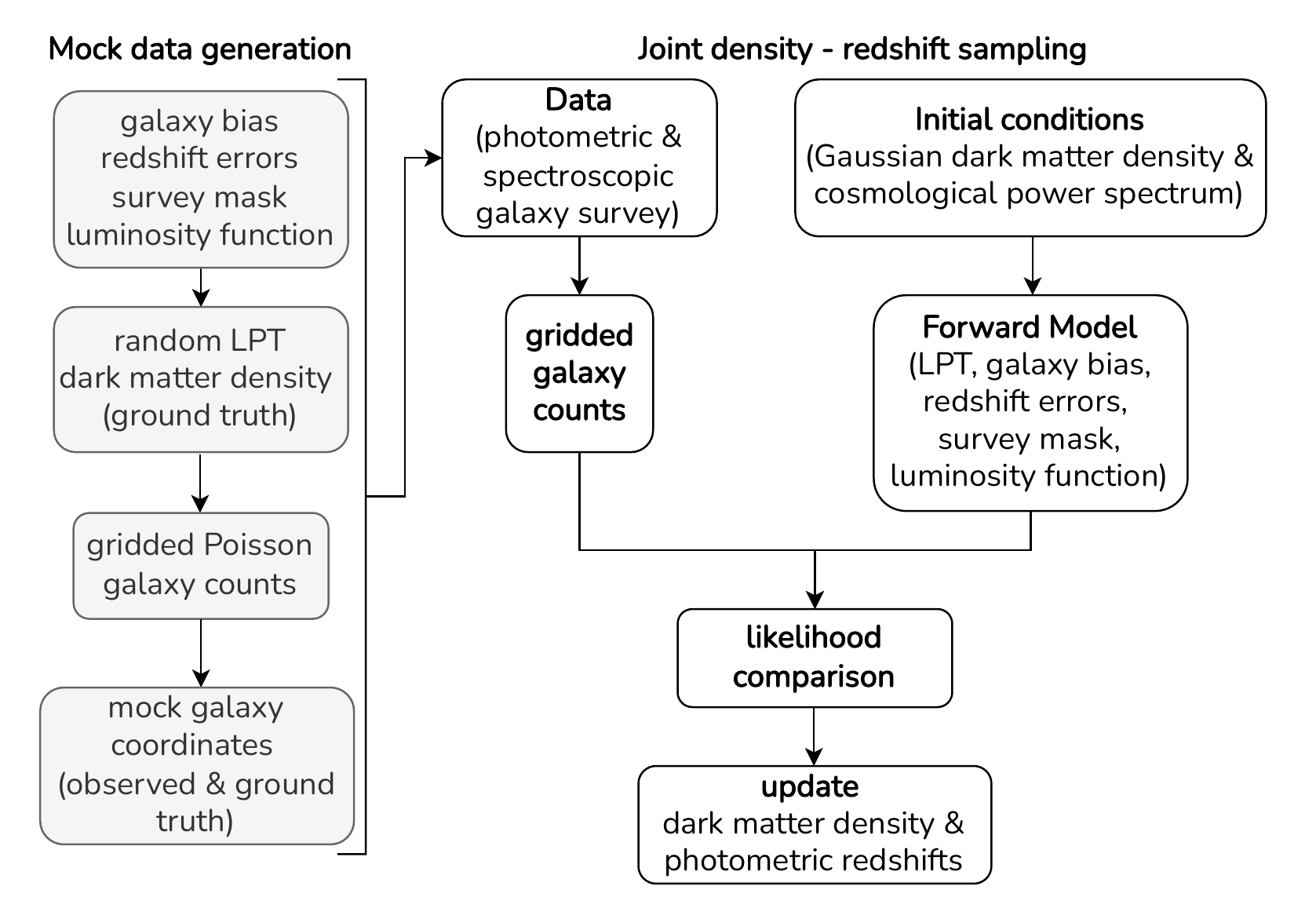}
    \caption{Flowchart of our algorithm. Left column: Mock data generation. The input is a luminosity function, a survey mask, galaxy bias parameters and redshift uncertainty. We generate a random LPT density field on which we apply galaxy bias. From that, we draw a Poisson galaxy count sample. We populate the 3D grid with mock observations given a survey mask and selection function. We finally apply Gaussian noise to the mock redshifts. This data is fed into our forward model. Right column: MCMC sampling framework. We then evolve them using Lagrangian Perturbation Theory and compare the output to the gridded galaxy counts. After the likelihood comparison, we sample the dark matter density jointly with photometric redshifts.}
    \label{fig:plot_1}
\end{figure*}

Synergies between spectroscopic and photometric surveys have been explored, intended mainly for photometric redshift calibration \citep[e.g.][]{2017ApJS..233...21R,2019arXiv190410439C}. In light of this, a multitude of techniques to improve redshift uncertainty has been proposed \citep[e.g.][]{2015APh....63...81N,2015ApJ...813...53M,10.1093/mnras/stx510,2017MNRAS.469.1186S,2019MNRAS.489.4817D,2020JCAP...12..001S,2020A&A...636A..90S,2020MNRAS.491.4768R,2021ApJS..256....9S,2022arXiv220707673L}. However, there are limitations to spectroscopy. Photometric surveys typically reach fainter magnitudes, higher redshifts, cover a larger portion of the color-space and have higher redshift completeness \citep[e.g.][]{2009arXiv0912.0201L,2010MNRAS.406..881B,Euclid:2013,LSST:2019}. In order to exploit these advantages and avoid biases in cosmological analyses, it is necessary to mitigate the basic limitation of photometry, low accuracy. \cite{2012MNRAS.425.1042J} demonstrated that clustering information from the large-scale structure alone can improve the accuracy of a purely photometric sample up to one order of magnitude. 

In this work, we developed a framework that, for the first time, infers the three-dimensional cosmic large-scale structure jointly with redshift probability density functions (PDFs) of photometric galaxies using a physical structure formation model. In this way, we improve upon photometric redshift uncertainty and self-consistently quantify observational errors in the large-scale structure inference. In doing so, we preserve all high-order statistics of the large-scale structure and use a physically-motivated structure growth model. 

Our prior is the homogeneity and isotropy assumption for the initial conditions of structure formation, as well as the physical model for the formation of the large-scale structure and Gaussian initial conditions. Our framework is built within the \codefont{Bayesian Origin Reconstruction from Galaxies (BORG)} algorithm \citep{Jasche:2013,2015JCAP...01..036J,Lavaux:2019,2019A&A...625A..64J}, which infers the initial conditions of structure formation from galaxy observations in a fully Bayesian approach. We include $1\%$ spectroscopic galaxies, similar to the expected galaxy number ratio between next-generation photometric and spectroscopic surveys \citep{2009arXiv0912.0201L,Euclid:2013,LSST:2019}. 

Control of the impact of photometric redshift uncertainties on large-scale structure inferences is a requirement for a wide range of studies \citep[see][for a review]{2019NatAs...3..212S}. First, this is relevant to peculiar velocity analyses \citep[e.g.][]{2008MNRAS.389L..47A,2012ApJ...751L..30H,2014MNRAS.444.3926J,2015MNRAS.447..132W,2016MNRAS.463.4083A,Mediavilla_2016,2020MNRAS.497.1275S,2021PhRvD.103j3507P,2022arXiv220703707T,2022arXiv220400023P}, which constrain the growth of structure, gravity and cosmological parameters  \citep[e.g.][]{1976ApJ...205..318P,2008MNRAS.389L..47A,2014MNRAS.444.3926J,2021PhRvD.103j3507P}. The evolution of the peculiar velocity divergence with redshift is predicted by $\Lambda$CDM \citep[e.g.][]{1980lssu.book.....P,1991ApJ...379....6N,1998PhRvD..58l4006M,2001CQGra..18.5115E,2020EPJC...80..757T,2021Ap&SS.366....4F} and can be used as a cosmological test. 

Accounting for photometric redshift uncertainties is crucial for weak lensing inferences \citep[e.g.][]{2008MNRAS.386..781M,2020A&A...637A.100W,2022MNRAS.509.3194P,2021MNRAS.502.3035P}. Moreover, photometric redshift uncertainties significantly affect estimates of intrinsic alignment \citep[e.g.][]{2007NJPh....9..444B,2015MNRAS.448.3391C,2022JCAP...08..003T,2022arXiv220701627F}, redshift-space distortions \citep[e.g.][]{1987MNRAS.227....1K,2011RSPTA.369.5058P}, Baryon Acoustic Oscillations \citep[e.g.][]{2015MNRAS.449..835R,2018MNRAS.477.3892C} and can hinder the disentanglement of dynamical dark energy from modified gravity \citep{2010MNRAS.409..737W}. Further, probing the large-scale structure with galaxy clustering at high-redshift through its gravitational potential can constrain galaxy formation and evolution \citep[e.g.][]{2018JCAP...07..030S,2022ApJ...924L...3T}. Moreover, it can constrain the scale of cosmic homogeneity, since next-generation surveys will be sensitive to large-scale clustering and therefore, to superclusters and voids \citep[e.g.][]{2014arXiv1403.5237B}. Last but not least, accurate inferences of the large-scale structure at high-redshift constrained by photometric galaxy clustering can be used complementary to Lyman-$\alpha$, which is sensitive to voids and quasar clustering, which are differently biased than regular galaxies \citep[e.g.][]{2014arXiv1403.5237B,2019A&A...630A.151P}.

The paper is structured as follows: In Section \ref{sec:data_model} we provide the mathematical formulation of the density and photometric redshift posteriors. In Section \ref{sec:method} we discuss the algorithmic implementation and configuration of the mock galaxy survey generator and the photometric redshift sampler. In Section \ref{sec:results} and \ref{sec:conclusions} we discuss our results and conclusions, respectively. 

\section{Statistical modelling of the large-scale structure}\label{sec:data_model}
We build our photometric redshift sampler on the \codefont{BORG} algorithm. \codefont{BORG} employs a hierarchical Bayesian forward-model to infer the posterior distribution of plausible initial conditions from which present structures have formed. To solve this statistical initial conditions problem, \codefont{BORG} uses a sophisticated Markov Chain Monte Carlo (MCMC) approach.  

The resulting large-scale structure posterior is approximated by realizations of the large-scale structure, constrained by galaxy observations. \codefont{BORG} takes into account the survey geometry, flux limitations and related systematic effects and accounts for them in the large-scale structure inference. In the present work, we constrain the primordial and present-day large-scale structure jointly with photometric galaxy observations, while quantifying all observational uncertainties.  

\subsection{Gibbs sampling of the joint posterior}

We achieve joint constraints on the large-scale structure and galaxy comoving distances by jointly sampling from their joint posterior. In order to do so, we choose a Gibbs sampling approach \citep{10.1093/biomet/57.1.97}, in which we first draw a real-space density sample conditioned on galaxy redshifts and then galaxy redshifts conditioned on the primordial density field, $\delta^i$
\begin{eqnarray}
    z_{j+1} &\curvearrowleft&   \mathcal{P}\left(z\given [\Big]z_\mathrm{obs},\theta,\delta^i_{j}\right)\label{eq:z_sample} \\
    \delta^i_{j+1} &\curvearrowleft& \mathcal{P}\left(\delta^i\given [\Big]z_{j+1}, z_\mathrm{obs},\theta\right)\label{eq:d_sample},
\end{eqnarray}
where $z$ is the sampled redshift of a galaxy, $z_\mathrm{obs}$ its observed redshift, $\theta$ the right ascension and declination and $j$ indicates the MCMC sampling step. We then explore the joint dark matter density and photometric redshift posterior by iteratively sampling from the above conditional distributions.

\subsection{Dark matter density}\label{density_posterior}
For the sake of demonstration we use first-order Lagrangian Perturbation Theory (LPT) to model structure formation and a Poisson likelihood to associate the dark matter density field to photometric galaxy observations. \codefont{BORG} allows for more complex structure formation models, such as the particle-mesh model \citep{2019A&A...625A..64J}, which can be used with the current photometric redshift sampling approach in the future. Within \codefont{BORG}, the galaxy coordinates are transformed to galaxy counts, $N^\mathrm{g}$, using Nearest Grid Point projection \citep{1974JCoPh..16..342E}. As we show in \refapp{dposterior}, \refeq{d_sample} can be written as $\P(\delta|\N)$. Below, we focus on the main aspect of our work, inferring the primordial density field, $\delta^i$, in conjunction with the late-time large-scale structure. We therefore write
\begin{equation}
    \P(\delta^i\given  N^\mathrm{g}) \propto \int d\delta \P(N^\mathrm{g}\given \delta)\P(\delta\given  \delta^i)\P(\delta^i),
    \label{eq:density_posterior}
\end{equation}
where $\delta^\mathrm{i}$ the primordial, real-space density field, $\P(N^\mathrm{g}|\delta)$ is the Poisson likelihood, $\P(\delta\given \delta^\mathrm{i})$ the structure formation term and $\P(\delta^i)$ the prior on the primordial density field. The Poisson likelihood is
\begin{equation}
    \mathcal{P}(N^\mathrm{g}\given \delta) = \prod_{k} \frac{\lambda{\strut_k}^{N^\mathrm{g}_k}e^{-\lambda \strut_k}}{N^\mathrm{g}_k!},
    \label{eq:Poisson}
\end{equation}
where $\lambda_k$ is the expected number of galaxies at the $k^\mathrm{th}$ grid element. The index $k$ runs over all grid elements. The dependence of the Poisson intensity, $\lambda_k$, on space models the inhomogeneous nature of the galaxy distribution. The Poisson intensity is given by
\begin{equation}
  \lambda_k=\mathcal{W}_k\langle n \rangle (1+\delta_k)^\beta,
    \label{eq:bias}
\end{equation}
where $\mathcal{W}_k$ is the survey window, $\langle n \rangle$ is the expected mean number of galaxies per grid element and $\beta$ the power-law exponent. For the sake of demonstration here, we have assumed a linear galaxy bias model, taking $\beta=1$. Our method can also account for nonlinear bias models, as demonstrated previously in \cite{2019A&A...625A..64J,Lavaux:2019,2020MNRAS.494...50C}. The survey window encapsulates information on the luminosity function through the radial completeness function, $\mathcal{C}$, and  survey mask, $\mathcal{M}$, as follows
\begin{equation}
   \mathcal{W}(\boldsymbol{x}) = \mathcal{C}(|\boldsymbol{x}|)   \mathcal{M}(\boldsymbol{\hat{n}}),
   \label{eq:window}
\end{equation}
$\boldsymbol{\hat{n}}$ being the unit vector along the line of sight to a galaxy located at $\boldsymbol{x}$. We write the structure formation term as
\begin{equation}
\P(\delta\given \delta^\mathrm{i})=\prod_k \delta^\mathrm{D}(\delta_k-F_k(\delta^\mathrm{i}))
\end{equation}
\noindent where $\delta^\mathrm{D}$ is the Dirac delta distribution, $k$ runs over the grid indices and $F_k$ is the structure formation model, for which we choose first-order LPT. The Dirac delta represents that we assume no error on our gravity model. We sample from \refeq{d_sample} using a Hamiltonian Monte Carlo sampler, as introduced in \codefont{BORG} by \cite{2010MNRAS.406...60J}. Finally, we assume that the initial conditions for structure formation are described by a zero-mean multivariate Gaussian distribution of the primordial dark matter density, $\delta^\mathrm{i}$
\begin{equation}
\P(\delta^\mathrm{i})= \frac{1}{{\sqrt{{|2\pi S|}}}}\exp\left({-\frac{1}{2}\sum_q^N\sum_r^N\delta^\mathrm{i}_qS_{qr}^{-1}\delta^\mathrm{i}_r}\right),
\end{equation}
where $|S|$ indicates the determinant of the covariance matrix, $S$, of the primordial density field.  

\subsection{Photometric redshift modelling}\label{PZ_posterior}

We write the conditional redshift posterior  that we use in our Gibbs sampling approach (see \refapp{zposterior}) for each individual galaxy, $i$, as
\begin{equation}
    \P(z_i\given {z_\mathrm{\mathrm{obs}}}_i,\delta) \propto \P(\delta\given  z_i)  \P(z_i\given {z_\mathrm{\mathrm{obs}}}_i)\mathcal{J}(z_i),
    \label{eq:redshift_posterior}
\end{equation}
where $z_i$ the true redshift of a galaxy $i$ and $\mathcal{J}$ the Jacobian matrix of the transform from redshift- to real-space volume element:
\begin{equation}
    \mathcal{J}(z_i) = \Bigg{|} r^2(z_i) \frac{\partial r}{\partial z}\Big{|}_{z_i} \Bigg{|}.
\end{equation}
We provide a detailed derivation of \refeq{redshift_posterior} in \refapp{zposterior}. The formulation in \refeq{redshift_posterior} is possible because each galaxy can be treated independently from all others given a density field, as a consequence of the Poisson density likelihood \citep{2012MNRAS.425.1042J}. The first term on the right-hand side is associated with the density field along the line of sight to the galaxy and the second term represents the photometric redshift likelihood. In real observations, photometric redshift likelihoods are highly non-Gaussian \citep[e.g.][]{2009MNRAS.400..429C}. Here, for demonstration, we choose a Gaussian likelihood with respect to the observed redshift, truncated at zero
\begin{equation}
    \P({z_\mathrm{\mathrm{obs}}}_i \given z_i) = \frac{T(z_i)}{\sqrt{2\pi\sigma^2(1+z_i)^2}} \exp{\left[-\frac{1}{2}\frac{({z_\mathrm{\mathrm{obs}}}_i-z_i)^2}{\sigma^2(1+z_i)^2}\right]},
    \label{eq:Gaussian}
\end{equation}
where
\begin{equation}
    T(z_i) = \left[\frac{1}{2}+\frac{1}{2}\mathrm{erf}{\left(\frac{z_i}{\sqrt{2}\sigma(1+z_i)}\right)}\right]^{-1},
\end{equation}
where $\mathrm{erf}(x)$ is the error function. This term ensures consistency with the truncation of observed redshifts at zero in the mock data. Our method generalizes to non-uniform photometric redshift uncertainties and can accept redshift estimates already constrained by other independent methods. Therefore, our algorithm can accommodate arbitrarily complex redshift distributions. Notice that we make the photometric redshift uncertainty dependent on redshift. We make this assumption as a proof-of-concept demonstration, but the method can account for any redshift likelihood. The Poisson term includes the radial component of the selection function, $\mathcal{C}_r$, and the line-of-sight density
\begin{equation}
\P(\delta\given  z_i) = \mathcal{C}_r(z_i)   \lambda(\boldsymbol{x}_i),
\end{equation}
in the range $0 < z_i \leq {z_\mathrm{max}}_i$, where ${z_\mathrm{max}}_i$ represents the maximum redshift that lies still within the observed volume. In order to draw redshift samples from \refeq{z_sample}, we use a slice sampler \citep{slice_sampler}. One can either update all galaxy redshifts at once, or one galaxy at a time. As demonstrated in \cite{2012MNRAS.425.1042J}, each galaxy can be treated independently given the density field. We therefore sample each galaxy individually and computationally in parallel.  In this process, galaxy redshifts at each sampling step are conditioned on the previous density field. Then, via \refeq{density_posterior}, the next density sample is conditioned on the updated redshifts. In this fashion, we  constrain photometric redshifts jointly with the dark matter density field. As a result, we reduce the uncertainty in both compared to inferring the density field from photometric redshifts directly.

\section{Mock data generation}\label{sec:method}

In this section, we describe the algorithmic implementation and configuration of the mock (ground truth) galaxy survey generator, as well as the photometric redshift sampler.

The input to our algorithm is right ascension, declination, observed redshifts, redshift uncertainty, survey mask, luminosity function and galaxy bias parameters. The output is a three-dimensional large-scale structure posterior and photometric galaxy PDFs, jointly constrained. The large-scale structure realizations are samples of the 3D large-scale structure posterior distribution and account for data- and survey-related uncertainties. Our framework goes beyond N-point correlations, as it infers the full 3D dark matter density posterior. Our method applies to regions where spectroscopic data coverage is not present or incomplete, because it can exploit information from photometric galaxy clustering alone. As a result, our method yields constraints also when using only photometric redshifts. Here we provide a proof-of-concept demonstration on mock data, focusing on the inference of cosmological fields jointly with photometric and spectroscopic observations with Gaussian redshift errors.

To validate and test the performance of our method we generate artificial photometric and spectroscopic surveys by the following procedure:
\begin{enumerate}
    \item Generate a random Gaussian primordial density field (initial conditions), $\delta^\mathrm{i}$ and the corresponding present-day density field, $\delta$ using the LPT forward model.
    \item Draw mock galaxy counts from a Poisson distribution conditional on the present-day density field
    \begin{equation}
         N^\mathrm{g}_\mathrm{true} \curvearrowleft   \P\left(N^\mathrm{g}_\mathrm{true}\Big{|}\delta\right),
    \end{equation}
    \item Here we displace galaxies in the volume elements. We start by drawing displacements for each galaxy from a uniform distribution, $\mathcal{U}$
    \begin{equation}
        u_{i,g} \curvearrowleft \mathcal{U}(0,1),
    \end{equation}
    where $i=(1,2,3)$ represents the three Cartesian directions, $g$ is the galaxy index, $u_i$ is the uniform displacement along direction $i$ for a given galaxy and $x_i$ is the galaxy's Cartesian coordinate $i$. We then displace galaxies in each cell of the three-dimensional grid as follows
    \begin{equation}
         x_{i,g} = {d_B}_{i} + R_{i}(n_{i,g} + u_{i,g} - 0.5),
    \end{equation}
    where ${d_B}_i$ is the $i$-coordinate of the lower left box corners and $n_i$ runs over the number of grid elements in one direction of the box. $R_i$ is the resolution along the direction $i$, defined as $R_i=L_i/N_i$, $N_i$ being the grid resolution along $i$ and $L_i$ the corresponding box size. The subtractive factor assigns galaxies to the lower left corner of each grid cell for the Nearest Grid Point projection.
    \item Iterate the following over galaxy counts in each grid cell
    \begin{enumerate}
        \item Transform Cartesian comoving coordinates to right ascension, declination, comoving distance, $r$ and redshift, $z$.
        \item Calculate the observation probability, $\mathcal{W}(\mathbf{x})$, at the galaxies' location, according to \refeq{window}. The observation probability depends on the luminosity function and  survey mask.
        \item Accept observed galaxies using rejection sampling: We draw a random number, $q$, in the range $(0,1)$. If $q<\mathcal{W}(\mathbf{x})$, we accept the galaxy and add it to the survey, otherwise we reject it. 
    \end{enumerate}
    \item We then generate observed redshifts by adding Gaussian noise with zero mean and variance $\sigma^2(1+z)^2$ to the ground truth redshifts, $z$, according to \refeq{Gaussian}. We truncate the observed redshifts at zero. 
    \item We generate an observed galaxy count field using Nearest Grid Point projection on the observed redshifts and a ground truth galaxy count field using the ground truth redshifts.
\end{enumerate}

We perform the inference in a box extending to redshift $z=0.8$, with a box size of $1660$\Mpc{}, a grid resolution of $128^3$ and a real-space resolution of $13$\Mpc{}. The observer is at the lower left corner of the box, such that the observed area covers $\sim 1$ octant of the sky. We assume the Planck 2018 cosmological parameters \citep{2020A&A...641A...6P}. We do not sample spectroscopic redshifts, as their uncertainties are insignificant compared to the resolution of our inference and photometric redshift uncertainties.

Our mock data consists of a catalog with $2\cdot10^7$ photometric and $2\cdot10^5$ spectroscopic galaxy redshifts, similar to the fraction of photometric to spectroscopic redshifts and number density in next-generation surveys in our redshift range \citep{2009arXiv0912.0201L,Euclid:2013,LSST:2019}. We take the photometric luminosity function to be an i-band Schechter with $M_* = -22.8$ and $\alpha = -1$ and the spectroscopic luminosity function to be a j-band Schechter with $M_* = -23.04$ and $\alpha = -1$ \citep[Table 2,][]{2012ApJ...752..113H}, as these are bands that will be used in next-generation surveys. The generation of the angular survey mask is described in \cite{2022arXiv220308838A}. The photometric and spectroscopic components of each catalog fully overlap. We adopt a worst-case scenario for photometric redshift uncertainties, $\sigma = 0.05(1+z)$ \citep{2009arXiv0912.0201L}. For illustrative purposes we choose a linear galaxy bias model. 


\begin{figure*}
    \centering
    \includegraphics[width=\textwidth]{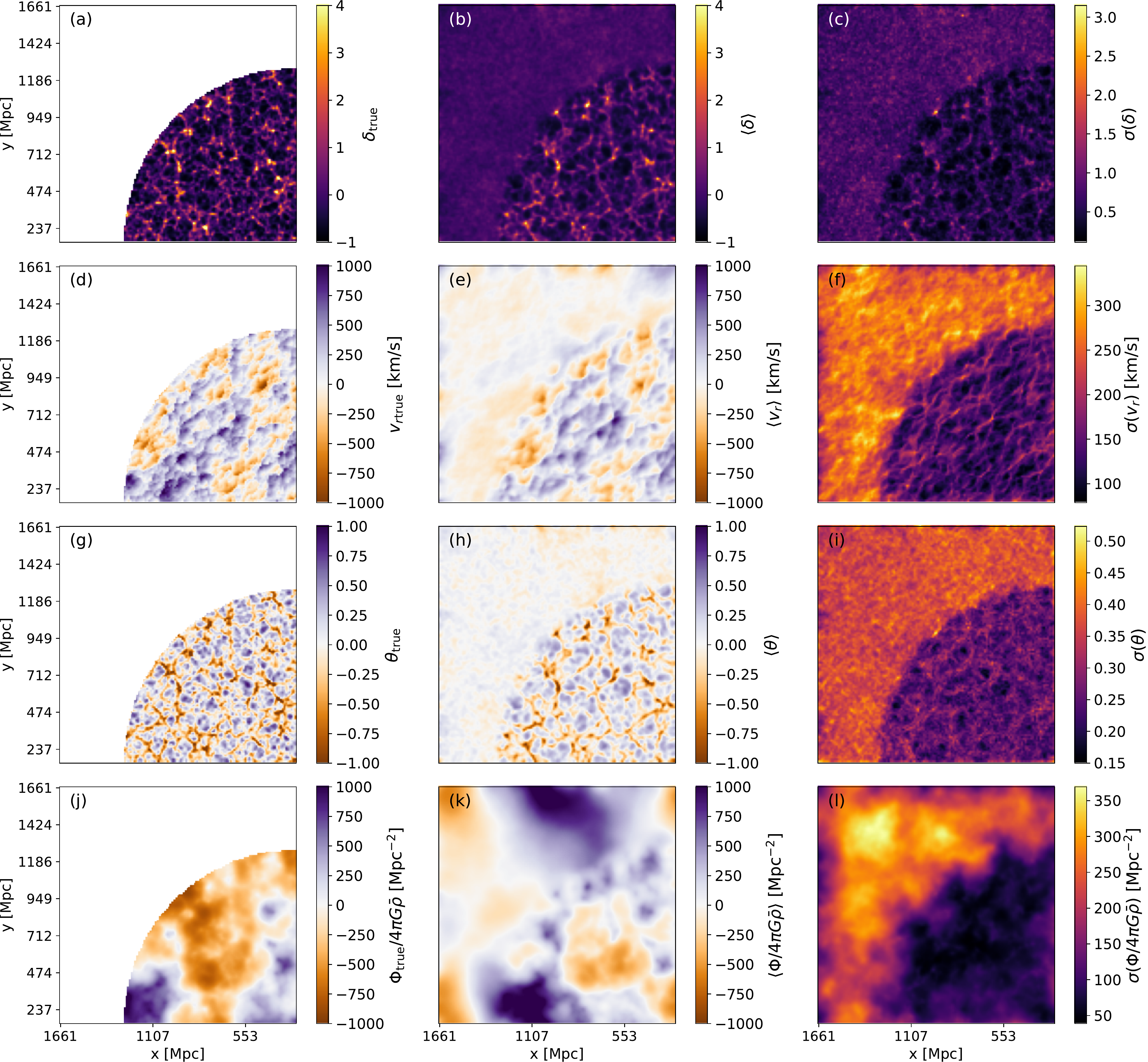}
    \caption{From top to bottom, slices through the three-dimensional dark matter density, radial peculiar velocity, divergence, gravitational potential and the off-diagonal components of the tidal shear for the ground truth (left column), average (middle column) and standard deviation (right column) across the MCMC samples. The slices are at the same fixed distance from the observer. The white region is outside the survey footprint. As shown in the right column, regions populated by galaxies have lower uncertainty than regions outside the survey footprint.}
    \label{fig:plot_4}
\end{figure*}

\begin{figure*}
    \centering
    \includegraphics[width=0.75\textwidth]{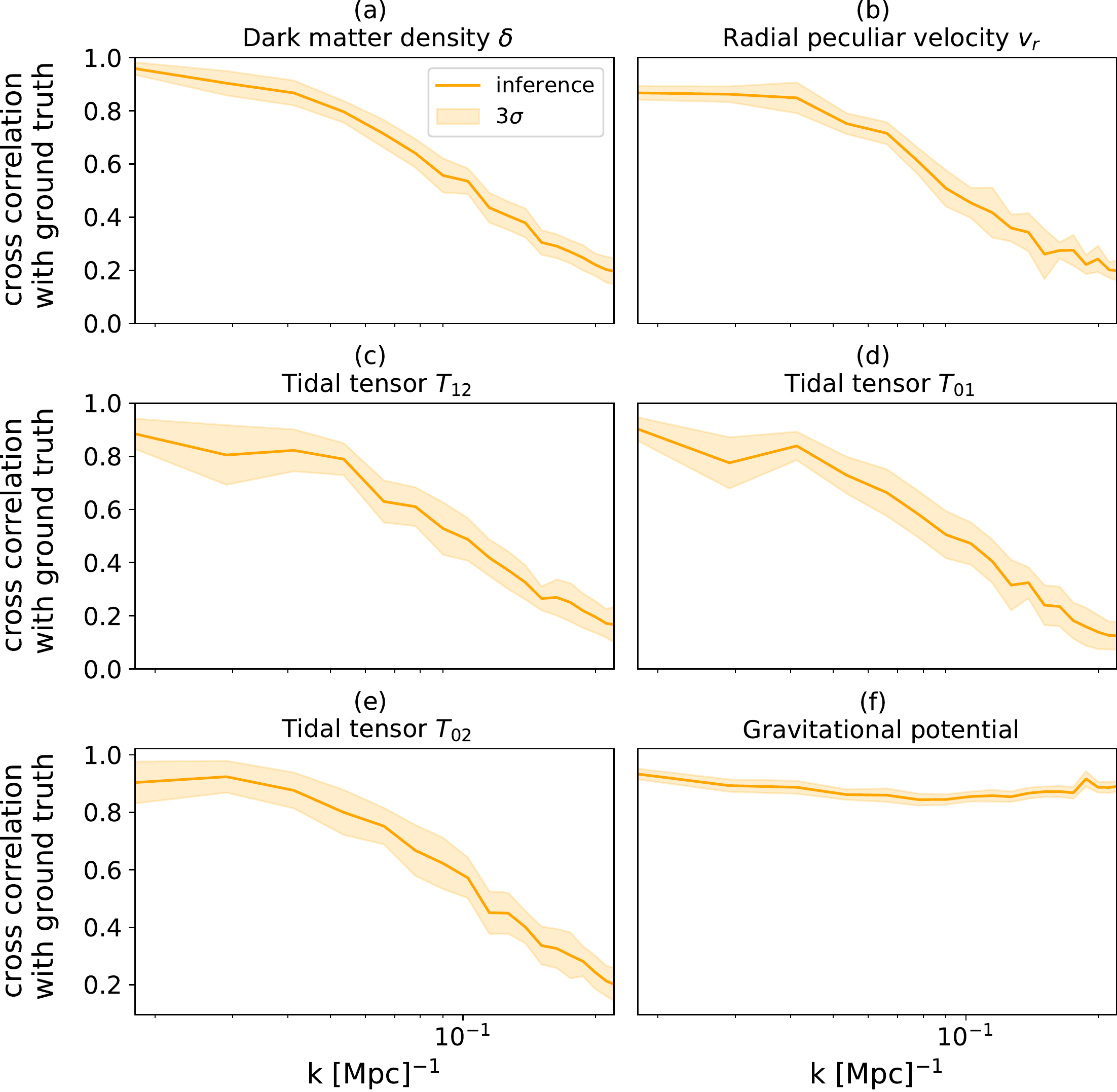}
    \caption{Cross-correlation of our inferred cosmological fields with the ground truth. The colored windows indicate 3$\sigma$ error bars. We consider a subbox that is the largest unmasked cubic volume in our inference domain, with side length $520$\Mpc{}. On the largest scales, we find a cross-correlation $>90\%$ with the ground truth for all cosmological fields. On the smallest scales, $\sim 0.04\langle \sigma_z \rangle$, we find a cross-correlation of $\sim 20 \%$ for the dark matter density, peculiar velocity and tidal tensor and $\sim 90\%$ for the gravitational potential.}
    \label{fig:plot_16}
\end{figure*}

\begin{figure*}
    \centering    
    \includegraphics[width=0.92\textwidth]{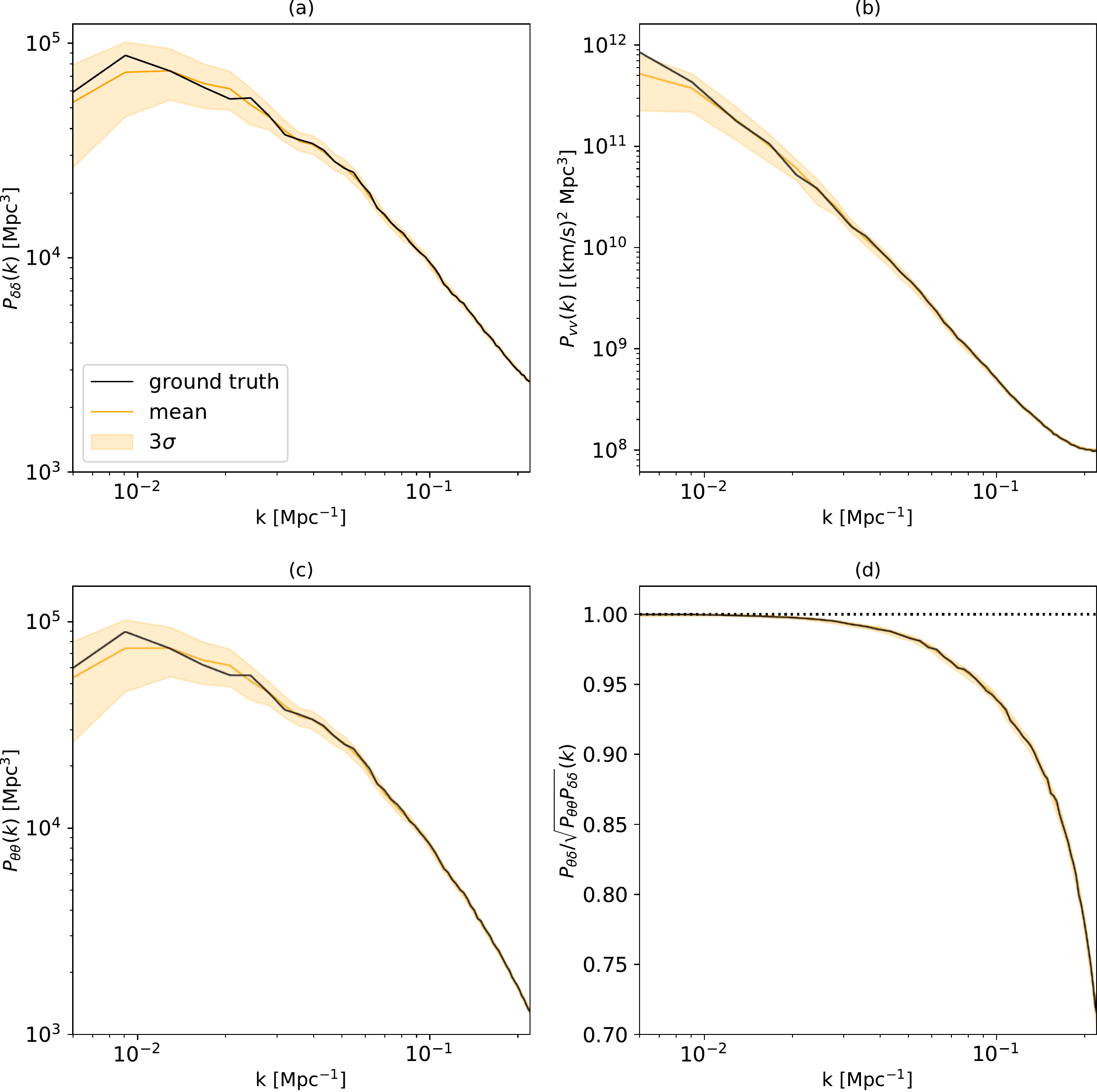}
    \caption{(a) Dark matter density, (b) radial peculiar velocity and (c) peculiar velocity divergence autocorrelation power spectrum. (d) Normalized cross-correlation between the dark matter density and peculiar velocity divergence. The black lines indicate the ground truth. The colored window represents $3\sigma$ error bars. The uncertainty on the 2-point statistics is due to galaxy survey- and data-related uncertainties. The blue line is the \codefont{CAMB} density autocorrelation power spectrum.}
    \label{fig:plot_5}
\end{figure*}

\begin{figure*}
    \centering
    \includegraphics[width=\textwidth]{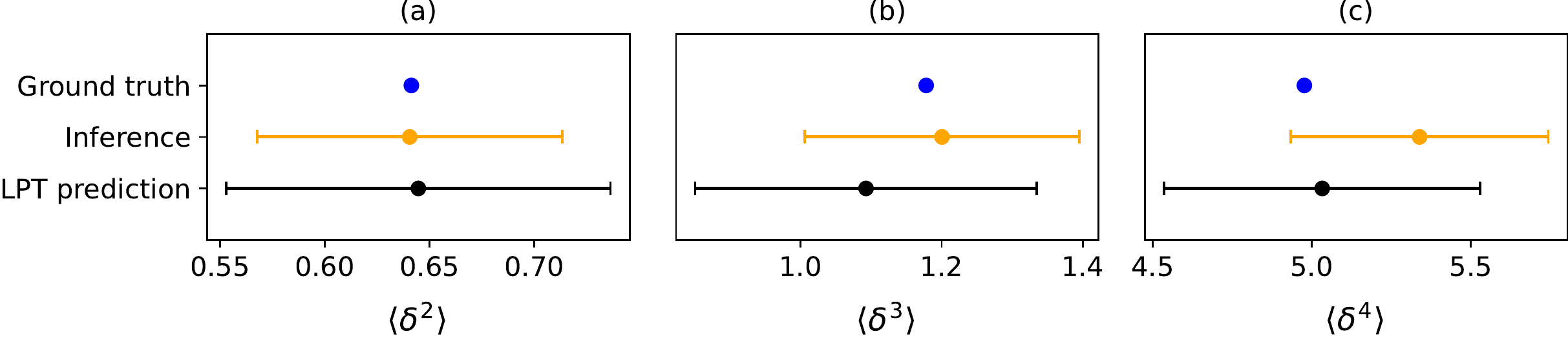}
    \caption{(a) Variance, (b) skewness and (c) kurtosis of the dark matter density field for our inference, the ground truth and a set of random LPT simulations at 13\Mpc{} ($\sim 0.04\langle \sigma_z \rangle$) using the entire inference domain. The latter two are random, unconstrained simulations. The inference is constrained jointly with galaxy observations. The error bars encapsulate the $1\sigma$ uncertainty both from the mock galaxy survey and the estimator. Our ground truth is consistent with a random LPT simulation and our inference is consistent with both. Our inference accurately captures higher-order statistics of the dark matter density field on scales much smaller than the original photometric redshift uncertainty.}
    \label{fig:plot_13}
\end{figure*}

\begin{figure*}
    \centering
    \includegraphics[width=\textwidth]{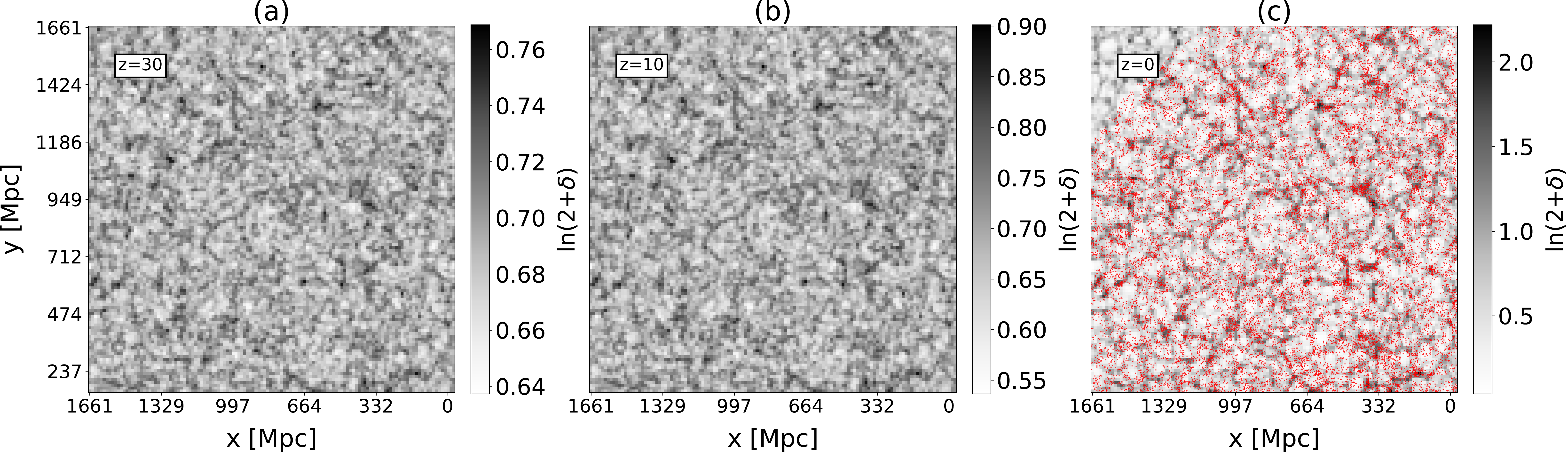}
    \caption{Structure formation history derived from the dark matter density field at (a) $z=30$, (b) $z=10$, (c) $z=0$ from a typical density sample. In red, we overlay the density slice with the galaxy coordinates in a typical realizations. Galaxies follow the present-day large-scale structure. The origins of high-density regions are already visible at $z=30$.}
    \label{fig:plot_8}
\end{figure*}

\begin{figure*}
    \centering
    \includegraphics[width=0.75\textwidth]{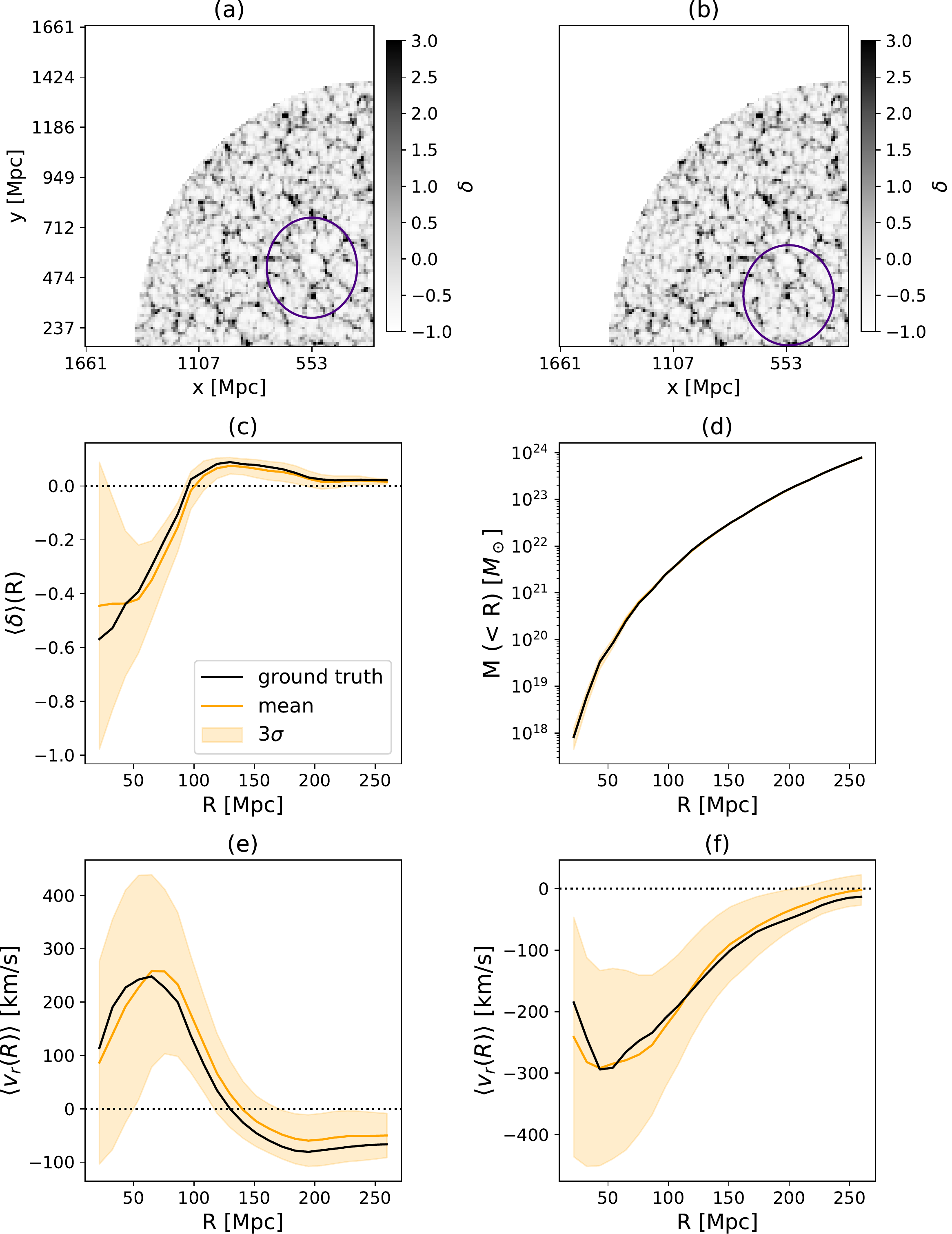}
    \caption{(a) Slice through the ground truth dark matter density field. In purple, a spherical shell around a low-density region. (b) Slice through the ground truth dark matter density field. In purple, a spherical shell around a high-density peak. (c) The mean density contrast in spherical shells around the low-density region. (d) The mass enclosed in shells around the high-density peak. (e) The mean radial peculiar velocity in spherical shells around the low-density region with respect to the center of the void. (f) The mean radial peculiar velocity in spherical shells around the high-density peak with respect to the center of the peak.}
    \label{fig:plot_9}
\end{figure*}

\section{Results}\label{sec:results}

\subsection{Constraints on the large-scale structure}

We present a qualitative illustration of our results in \reffig{plot_2}. In \reffig{plot_2}a we show the observed galaxy positions in our mock survey that are radially-distorted due to Gaussian redshift noise. In comparing that to the galaxy coordinates as constrained by our algorithm in \reffig{plot_2}b, we see that despite the initially large redshift uncertainty, photometric galaxies now trace filamentary structures in the dark matter distribution. These positions are derived from a typical MCMC sample. Observational uncertainties are accounted for in the entire set of MCMC samples. We show the ground truth galaxy positions in \reffig{plot_2}c. We notice high visual resemblance between the constrained observations and the ground truth. In \reffig{plot_3} we show the cross-correlation between the gridded photometric galaxy coordinates with the ground truth before and after the application of our method. We select an unmasked subvolume of our inference domain to demonstrate the best-case improvement in cross-correlation given the uncertainties in our survey configuration. We find the largest improvement in the cross-correlation to be on scales $\sim 0.5 \langle \sigma_z \rangle$, from $28\%$ to $86\%$. This indicates that our method provides both accurate inferences of the large-scale structure and reduces photometric redshift uncertainty. We refer the reader to \reffig{plot_1} for the flowchart of our method.

In \refapp{data} we provide a description of the estimators we use to derive properties of the large-scale structure, along with scientific cases that call for use of the derived properties. In \reffig{plot_4} we show statistical summaries of the posterior distribution of cosmological fields. In \reffig{plot_4}a we show the ground truth dark matter density field. The white region represents unobserved regions outside the survey footprint. As \codefont{BORG} effectively returns an unconstrained simulation in this region, we remove it for easier comparison to the other plots. In \reffig{plot_4}b, we show the ensemble mean density across the MCMC realizations. In the volume populated by galaxies the similarity between the mean and the ground truth is visible. Outside this volume, the ensemble mean tends to the cosmic mean. This is expected for an ensemble of random density fields, because there is no galaxy clustering information. Further, notice that overdense structures are smeared in the inference mean. Galaxies in each MCMC sample move along their line of sight. Therefore, the smearing is due to averaging over density realizations that account for photometric redshift uncertainties. In \reffig{plot_4}c we show the voxel-wise standard deviation of the dark matter density field. This includes observational uncertainties and shot noise due to the finite number of galaxies in each voxel. 

In \reffig{plot_4}d-f, we present a slice through the peculiar velocity field. In \reffig{plot_4}d we show the ground truth radial peculiar velocity field, derived from the ground truth dark matter density field with the dark matter sheet estimator. In \reffig{plot_4}e we show the radial peculiar velocity mean across the MCMC realizations. In \reffig{plot_4}f we present the sample variance of the radial peculiar velocity posterior. In \reffig{plot_4}g-i, we show the radial peculiar velocity divergence ground truth, ensemble mean and standard deviation. 

In \reffig{plot_4}j-l, we show our constraints on the gravitational potential from photometric and spectroscopic galaxy clustering. In \reffig{plot_4}j we show the ground truth gravitational potential, as derived from the ground truth dark matter density field. In \reffig{plot_4}k and  \reffig{plot_4}l, we show the ensemble mean and standard deviation across the MCMC realizations, respectively.

The above statistical summaries present high visual resemblance with the ground truth. We quantify this resemblance in \reffig{plot_16}, by showing the cross-correlation of the dark matter density, radial peculiar velocity, off-diagonal components of the tidal shear tensor and gravitational potential with the ground truth. We select an unmasked subvolume of the inference domain, to showcase the best-case cross-correlation in regions with data. On the largest scales, as expected, we have the highest cross-correlation for all cosmological fields. The cross-correlation of the gravitational potential with the ground truth is $>80\%$ on scales, as $\Phi$ has the longest correlation length. 

In \reffig{plot_5} we show the autocorrelation power spectra of cosmological fields that are typically used in cosmological analyses. ll power spectra were estimated and corrected for the \codefont{BORG} Cloud-In-Cell mass assignment scheme using \codefont{Pylians} \citep{Pylians}, as our density field has been estimated with a Cloud-In-Cell estimator. The autocorrelation power spectrum, $P_{\delta\delta}$, is shown in \reffig{plot_5}a. The two-point statistics of the inferred density fields are consistent with the ground truth and the sampler has covered the uncertainty due to cosmic variance \citep[e.g.][Figure 2]{2010PhRvD..81j4023P}. In \reffig{plot_5}b, we show the autocorrelation power spectrum of the radial peculiar velocity field, $P_{vv}$. The inference is consistent with the ground truth. 

In \reffig{plot_5}c, we show the peculiar velocity divergence autocorrelation power spectrum, $P_{\theta\theta}$, which is also consistent with the ground truth and $\Lambda$CDM prediction \citep[e.g.][]{2017MNRAS.467.3993A}. Further, we see that the power of $P_{\theta\theta}$ is suppressed compared to $P_{\delta\delta}$ beyond $k\sim0.05$ Mpc$^{-1}$. This is expected because collapsing structures that have virialized experience less volume change \citep[e.g.][]{2017MNRAS.467.3993A}. As a result, the peculiar velocity divergence field evolves less than the density field \citep[e.g.][]{2012MNRAS.425.2422K,2012MNRAS.427L..25J,2015MNRAS.454.3920H,2017MNRAS.467.3993A}. In \reffig{plot_5}d, we show the normalized cross-correlation power spectrum between the velocity divergence and dark matter density. On large scales, it is expected that to linear order $\theta \propto \delta$ \citep[e.g.][]{2015MNRAS.454.3920H}. On smaller scales, structure formation becomes nonlinear, which LPT is able to capture. A more refined gravity model, like a particle mesh \citep[][in \codefont{BORG}]{2019A&A...625A..64J}, would be needed to push these results to smaller scales and capture the generation of peculiar velocity vorticity.

In \reffig{plot_13}, we show higher-order moments of the constrained dark matter density field for the ground truth, our inference and the LPT prediction, with 1$\sigma$ error bars at the target resolution of 13\Mpc{} ($\sim 0.04 \langle \sigma_z \rangle$). We derive the LPT prediction from a set of random simulations ran using the same cosmological parameters as our main inference, but without constraints from galaxy redshifts. The error bars naturally account for observational uncertainties and the estimator uncertainty related to the sample size \citep{TQMP10-2-107}. For this fully self-consistent setting, our results are consistent with the ground truth and LPT prediction within 1$\sigma$. This result suggests that we infer the density field, including its higher-order moments, on scales much smaller than the original photometric redshift uncertainty.

In \reffig{plot_8} we show slices through the same randomly-selected realization in the structure formation history. Under the assumption of a causal structure formation model, we reconstruct the structure formation history which gives rise to the observed structures today. The initial conditions are set at $z=99$, but we show slices up to $z=30$, such that the seeds of present-day structures are visible. On the $z=0$ slice we overlay the ground truth locations of galaxies. Galaxies closely trace clusters and filamentary structures, whereas voids are sparsely populated by galaxies. This is because the Poisson noise is lower in higher-density regions. As expected, the origins of high-density peaks in the dark matter density field are already visible at early times. 

In \reffig{plot_9}a, we show a void in the volume covered by observations. In \reffig{plot_9}c we show the average density contrast in shells around the void. We see that the density contrast tends to zero, as expected when averaging over a scale close to the cosmic homogeneity scale \citep[e.g.][]{2018MNRAS.475L..20G}. In \reffig{plot_9}b, we show the location of a cluster in the density field. In \reffig{plot_9}d we show the mass enclosed in shells around the cluster. We derive it using the prescription in \cite{2019A&A...630A.151P}. In this study, the particle mass is $3.45\times10^{13}\;\mathrm{M}_\odot$. Overall, these results suggest that we recover correctly both the statistical properties of the large-scale structure around voids and clusters, but also individual structures.

We further show radial peculiar velocity profiles around the same void and cluster. The radial peculiar velocity becomes positive close to the void, indicating matter flowing outside the shells and toward overdense regions, as expected from gravitational structure growth \citep{1980lssu.book.....P}. On larger scales the average peculiar velocity tends to zero due to cosmic homogeneity. In \reffig{plot_9}f we show the average radial peculiar velocity in shells of radius $R$ around the cluster in \reffig{plot_9}b, with respect to its center. The radial peculiar velocity is negative close to the cluster, as expected from gravitational infall. On larger scales, the average velocity tends to zero as expected from cosmic homogeneity. 

\subsection{Constraints on photometric redshifts}

\begin{figure*}
    \centering    
    \includegraphics[width=\textwidth]{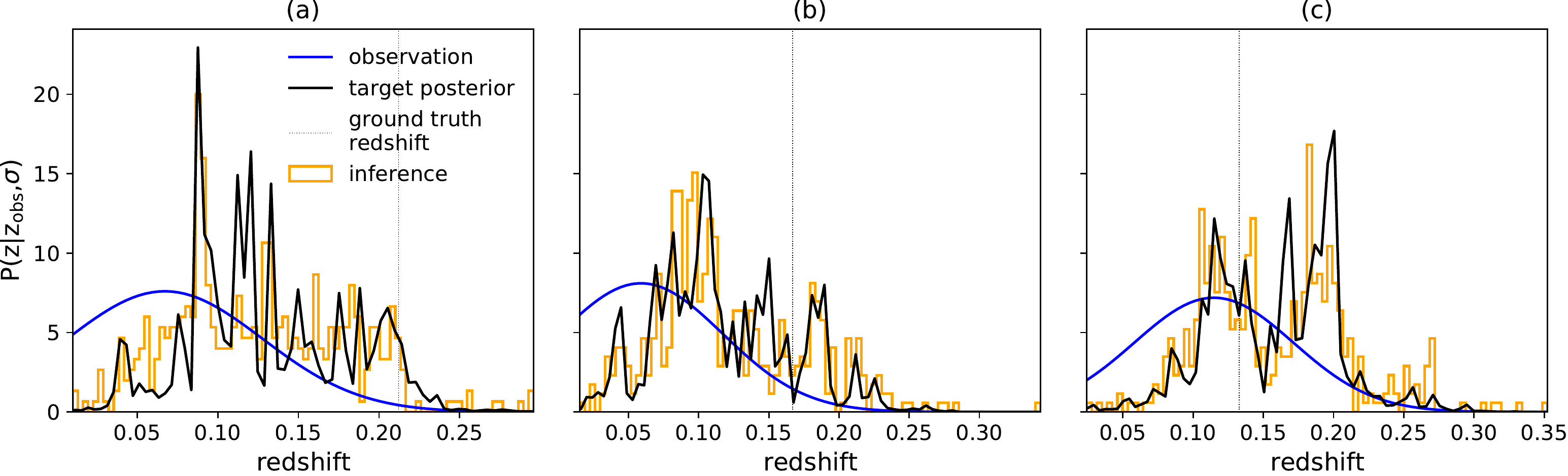}
    \caption{Photometric redshift samples for three randomly-selected galaxies. Overlayed are the corresponding initial observations estimated from our pipeline without a physical prior, the ideal photometric redshift posterior (for physical prior, using the ground truth density) and the ground truth redshift. The maximum redshift to which the posteriors extend indicates where each galaxy's line of sight intercepts the inference box. The inferred posterior is the marginal over density realizations, accounting for observational and survey-related uncertainties. The redshift binning is determined by the real-space resolution of the inference.}
    \label{fig:plot_11}
\end{figure*}

In this section, we give a brief overview of how the joint inference of the large-scale structure and photometric redshift PDFs yields constraints on photometric redshifts. In \reffig{plot_11} we show our results for three randomly-selected galaxies. The redshift likelihood is a Gaussian PDF. Its mean is the mock observed redshift and its standard deviation is $0.05(1+z)$, $z$ being the mock cosmological redshift. We derive the target posterior by multiplying the Gaussian likelihood with the Poisson term for the redshift inference in our Gibbs sampling approach, as shown in \refeq{redshift_posterior}. The results of our inference are overlayed as a histogram. This qualitative demonstration illustrates that the photometric redshift posterior after the application of our method contains more information than the likelihood and follows the radial density profile of the large-scale structure along the line of sight to each galaxy. This mechanism constrains the radial positions of galaxies from the initial, radially-distorted ones (\reffig{plot_2}a), to the final ones, that trace the filamentary large-scale structure (\reffig{plot_2}b). 

\begin{figure}
    \centering
    \includegraphics[width=8cm]{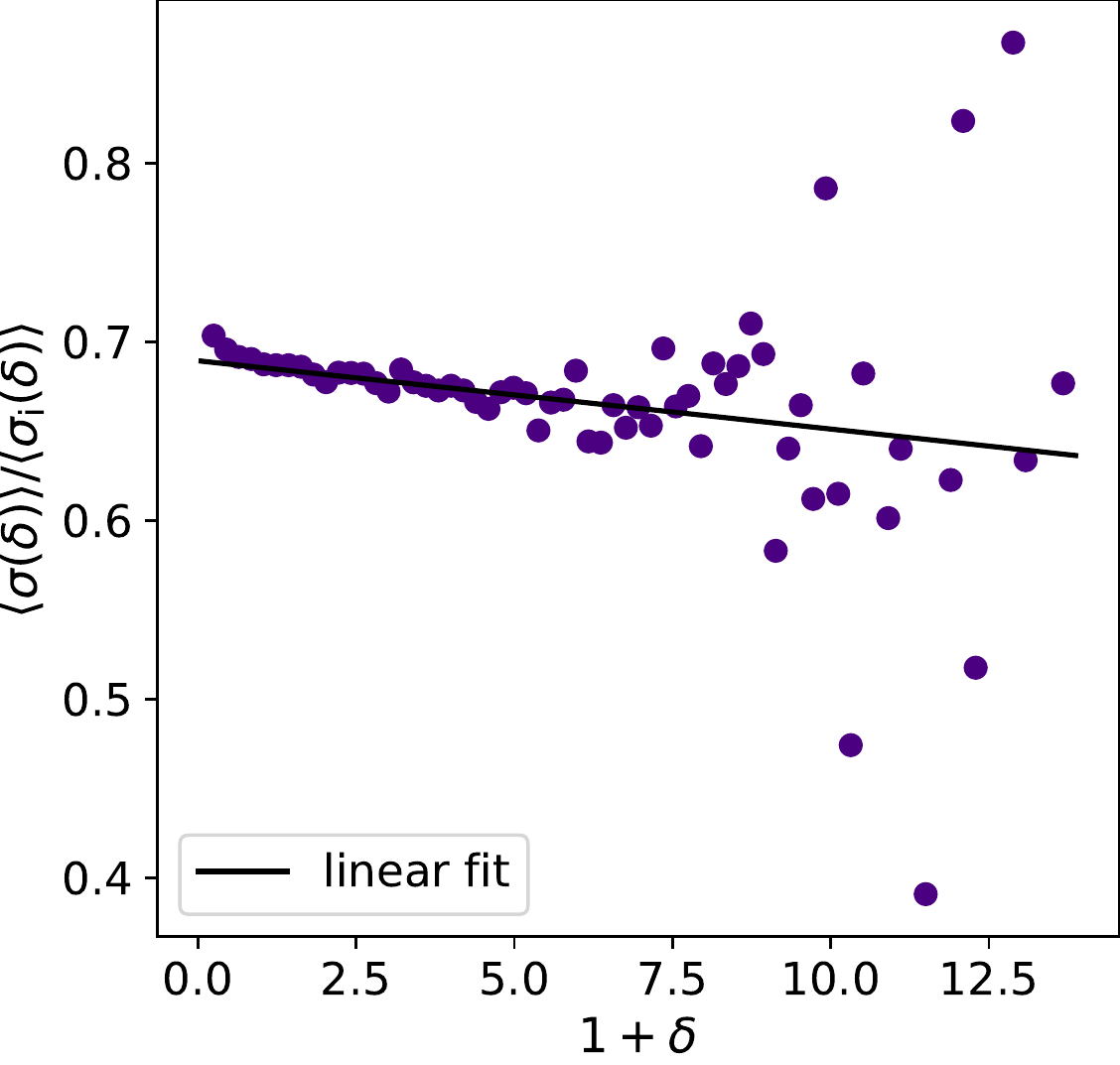}
    \caption{Final to initial photometric redshift uncertainty as a function of density at the ground truth galaxy locations. The black line is a linear fit, indicating that the reduction in redshift uncertainty is greater in regions of high dark matter density, as expected from the Poisson likelihood. The dispersion on the high-density end is due to shot noise, as there are fewer high-density peaks in the inference domain.}
    \label{fig:plot_15}
\end{figure}

\begin{figure}
    \centering
    \includegraphics[width=8.5cm]{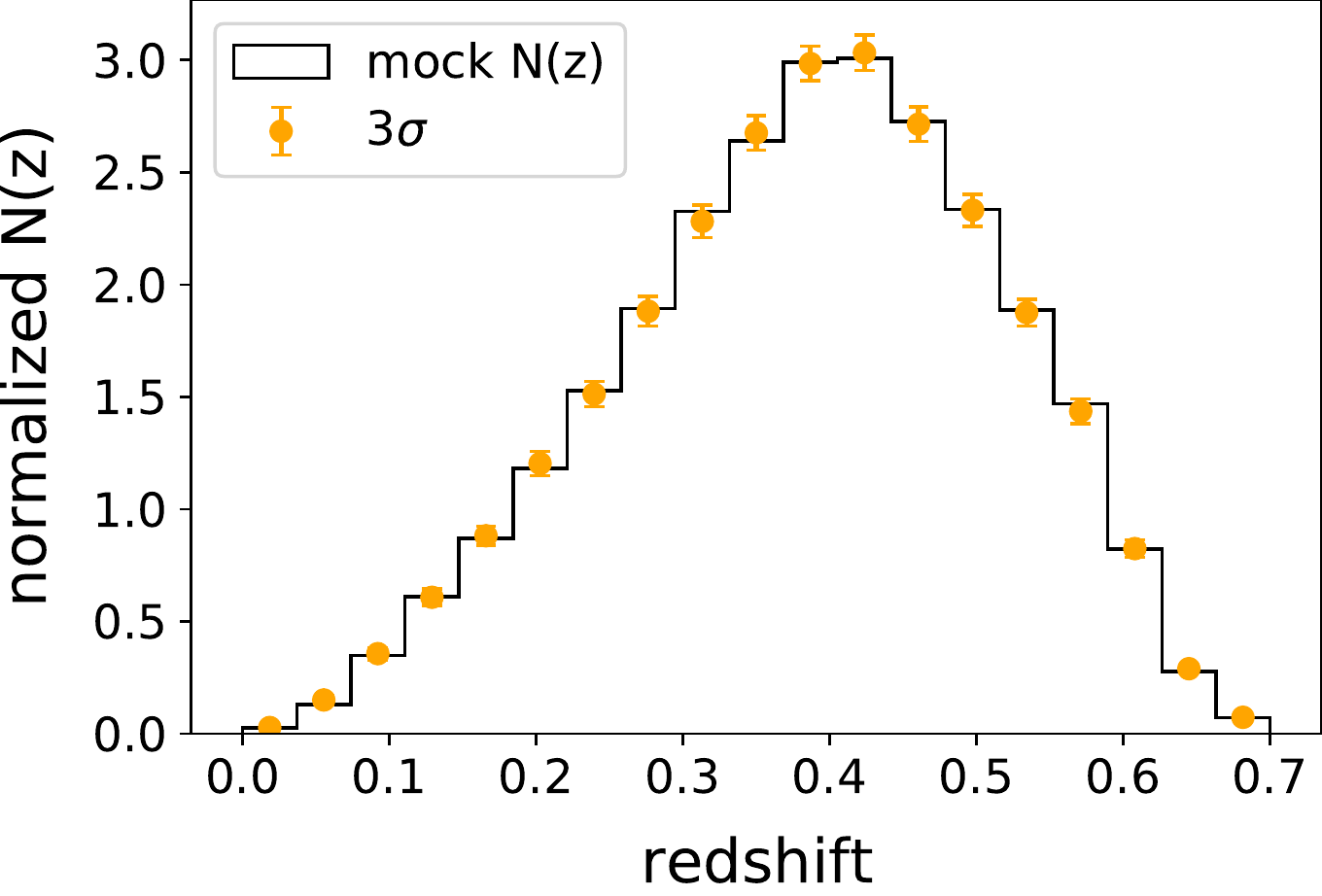}
    \caption{In orange, the average of normalized N(z) of inferred photometric redshifts across MCMC samples for a subsample of galaxies, along with $3\sigma$ uncertainties. In black, the mock photometric N(z) for the same subsample. The N(z) of inferred photometric redshifts is consistent with the ground truth N(z).}
    \label{fig:plot_17}
\end{figure}

In \reffig{plot_15} we show the reduction in the mean standard deviation of the photometric redshift posteriors as a function of density after the application of our algorithm. $\sigma_i$ indicates the original photometric redshift uncertainty of the mock observations. We expect that higher-density regions yield better constraints and hence, lower redshift uncertainty. The linear fit suggests that $\langle \sigma(\delta) \rangle / \langle \sigma_i(\delta) \rangle=-0.005(1+\delta)- 0.693$. In low-density regions, the mean standard deviation reduces by a factor of $0.7\langle \sigma_i \rangle$, whereas in high-density regions the reduction reaches $0.4\langle \sigma_i \rangle$. The scatter is larger in high-density regions because there are fewer strongly overdense peaks. Overall, we expect the redshift uncertainty reduction to be more significant in higher-resolution settings, where higher-density peaks will be resolved. It should be noted, however, that due to the highly multimodal nature of the target redshift distribution, the reduction in standard deviation does not reflect the information gain from the original to the final redshift PDF. Finally, in \reffig{plot_17} we compare the inferred distribution of photometric redshifts, $N(z)$, in our mock survey to the ground truth $N(z)$. For legibility, we show the mean $N(z)$ across the MCMC samples, along with $3\sigma$ error bars. The inferred $N(z)$ is consistent with the ground truth $N(z)$. We postpone the sampling of the $N(z)$ to future work. 

\section{Conclusions}\label{sec:conclusions}

Next-generation photometric galaxy surveys will deliver an extraordinary amount of observations, that will reduce the observational uncertainties, will extend deeper in redshift and cover a larger cosmological volume than spectroscopic ones. At the same time, most cosmological information is in the smallest cosmological scales which cannot be accessed in presence of large photometric redshift uncertainties. In such a setting, it is paramount to obtain control of photometric redshift uncertainties in cosmological inferences. Accurate inferences of the large-scale structure constrained jointly with photometric redshifts offer this possibility.

In this study we developed a method to constrain the primordial and present-day cosmic large-scale structure jointly with photometric redshifts at a resolution of 13\Mpc{} using, for the first time, a structure formation model in a Bayesian forward modelling approach. We achieved these joint constraints through Bayesian inference of the initial conditions of structure formation with photometric galaxy clustering. Our method takes into account data- and survey-related uncertainties and preserves all higher-order statistics of cosmological fields. In this study, we used first-order LPT, yet our algorithm can incorporate any gravitational model \citep[e.g.][]{2015JCAP...01..036J,2019A&A...625A..64J}. We demonstrated our constraints using a fully overlapping mock photometric and spectroscopic galaxy catalog as a proof-of-concept. We assumed a worst-case scenario for photometric redshift uncertainties in stage-IV surveys, $\sigma_z = 0.05(1+z)$, and demonstrated that our method can reduce this uncertainty. 

In particular, we showed large improvement in the cross-correlation of the photometric galaxy positions with the ground truth. The maximum improvement was on scales $\sim 0.5 \langle \sigma_z \rangle$, where the cross-correlation increased from $28\%$ (in the mock photometric observations) to $86\%$ (in our inference). We further achieved accurate inferences of the dark matter density, peculiar velocities, gravitational potential and tidal shear on scales $\sim 0.04 \langle \sigma_z \rangle$.

We demonstrated the ability of our method to accurately capture the 2-point statistics of the dark matter density, as well as its skewness and kurtosis on scales $\sim 0.04  \langle \sigma_z \rangle$. Higher-order statistics have been promising in constraining dark energy \citep{2020PhRvD.101b3518V, 2022PhRvD.105j3521F}. As we can use our method with a particle mesh, our joint inference framework can be used to constrain the bispectrum with photometric galaxy clustering. 

Voids have been extensively used to probe structure formation close to the linear regime \citep[e.g.][]{2019MNRAS.490.4907D,2019BAAS...51c..40P,2021MNRAS.507.2267D,2021MNRAS.500.4173S,2022A&A...667A.162C}, however photometric uncertainties limit their detection \citep[e.g.][]{2017MNRAS.465..746S}. For this purpose, we explored the possibility of detecting cosmic structures much smaller than the original photometric redshift uncertainty. The galaxy positions that were originally radially smeared due to the presence of photometric uncertainties, accurately trace the filamentary pattern of the large-scale structure after the application of our method. Further, we are able to accurately capture individual structures, like voids and clusters. This is a crucial advantage of embedding a structure formation model in our forward model. Overall, we demonstrated that we accurately recover the statistical properties of the large-scale structure on scales much smaller than the original photometric redshift uncertainty. 

The present work opens up the possibility to mitigate photometric redshift uncertainties in next-generation photometric surveys. Concurrently, the incorporation of a structure formation model paves a new way forward to extract as much information as possible from the smallest cosmological scales, while going beyond 2-point statistics and preserving information in the data. 

\section*{Data and software availability}

Data products can be made available upon reasonable request.

\section*{Author Contributions}
The main roles of the authors were, using the CRediT (Contribution Roles Taxonomy) system (\url{https://authorservices.wiley.com/author-resources/Journal-Authors/open-access/credit.html}):\\
\textbf{E.T.:} conceptualization, methodology, software, formal analysis, validation, writing - original draft; \textbf{J.J.:} conceptualization, methodology, software (supporting, transferred \cite{2012MNRAS.425.1042J} to new \codefont{BORG} version), supervision, resources, writing - feedback, funding acquisition; \textbf{G.L.:} software (supporting, debugging and notably during the Message Passing Interface parallelization), writing - feedback, funding acquisition; \textbf{F.L.:} software (supporting, author of the Simplex-In-Cell estimator), writing - feedback.

\section*{Acknowledgements}

ET thanks Metin Ata and Natalia Porqueres for helpful discussions and feedback, and Adam Andrews, Deaglan Bartlett, Andrija Kostic, James Prideaux-Ghee, Fabian Schmidt, Ben Wandelt and Pauline Zarrouk for comments on the original manuscript. Part of the computation and data processing in this study were enabled by resources provided by the Swedish National Infrastructure for Computing (SNIC) at Tetralith, partially funded by the Swedish Research Council through grant agreement no. 2020-05143. This research utilized the HPC facility supported by the Technical Division at the Department of Physics, Stockholm University. This work was enabled by the research project grant ‘Understanding the Dynamic Universe’ funded by the Knut and Alice Wallenberg Foundation under Dnr KAW 2018.0067. JJ acknowledges support by the Swedish Research Council (VR) under the project 2020-05143 -- "Deciphering the Dynamics of Cosmic Structure". GL acknowledges support by the ANR BIG4 project, grant ANR-16-CE23-0002 of the French Agence Nationale de la Recherche, and the grant GCEuclid from "Centre National d’Etudes Spatiales" (CNES). This work was supported by the Simons Collaboration on ``Learning the Universe''. This work is conducted within the Aquila Consortium (\url{https://aquila-consortium.org}). 

\begin{appendices}

\section{Derivation of the conditional density posterior for real-space inference}\label{app:dposterior}

In this section, we describe how to arrive at the conditional density posterior of \refeq{density_posterior} starting from \refeq{d_sample}. The former equation shows how we incorporate a gravitational structure growth model into our framework, whereas the latter represents the output of the Gibbs sampling process. Here, we discuss how we obtain constraints on the real-space late-time density field using galaxy observations in redshift space.

The present-day density field is conditionally independent of the observed galaxy redshifts given an ensemble of sampled redshifts. Therefore
\begin{eqnarray}    \P(\delta|z,z_\mathrm{obs},\theta)&=&\P(\delta|z,\theta)\nonumber\\
    &=& \sum_{\N} \P(\delta,\N|z,\theta)\nonumber\\
    &=& \sum_{\N} \P(\delta|\N)P(\N|z,\theta).
\end{eqnarray}
The last term in the sum indicates how we arrive from redshift-space observations to real-space galaxy counts. As elaborated on in \cite[][Eq. B4]{2012MNRAS.425.1042J}, this involves a transform from redshift- to real-space such that
\begin{equation}
    \P(\delta|z,\theta)=\sum_{\N} \P(\delta|\N)\P(\N|\mathbf{x}(z,\theta)),
    \label{eq:conditional_d}
\end{equation}
$\mathbf{x}$ being the real-space galaxy positions. In what follows, we will omit dependence on $(z,\theta)$ for legibility. The latter term indicates how we bin galaxies onto a grid given their real-space positions. For this gridding operation, we use Nearest Grid Point projection such that
\begin{equation}
\P(\N|\mathbf{x})=
\prod_i \delta^D \left(\N_i - \sum_p W_\mathrm{NGP}(\mathbf{x}_i-\mathbf{x}_p)\right),
\end{equation}
$i$ being the voxel index, $p$ being the galaxy index and
\begin{equation}
W_\mathrm{NGP}(\mathbf{x}) = \prod_{n=1}^3 \begin{cases} 1 & \text{if } |x_n|N_n/L_n < 1 \\
0 & \text{otherwise,}      
\end{cases}
\end{equation}
Substituting the above into \refeq{conditional_d} we arrive at  
\begin{equation}
\P(\delta|z,\theta) = \P(\delta|\N).
\end{equation}

\section{Derivation of the conditional redshift posterior for real-space inference}\label{app:zposterior}

In this subsection we describe how we arrive at the conditional redshift posterior in \refeq{redshift_posterior}, while performing an inference of the density field in real space. Let us denote by $\theta$ the right ascension and declination of a galaxy and by $u$ the redshifts of all other galaxies in the previous sampling step. We start by the redshift posterior of a given galaxy $i$ and apply Bayes' law
\begin{eqnarray}
\P(z_i|\theta, u, \delta, z_{\mathrm{obs}_i}) &=& \P(z_i)\frac{\P(\theta, u, \delta, z_{\mathrm{obs}_i},z_i)}{\P(\theta, u, \delta, z_{\mathrm{obs}_i})}\nonumber\\
&=&\P(z_i)\frac{\P(z_{\mathrm{obs}_i}|\theta, u, \delta,z_i)\P(\theta,u,\delta|z_i)}{\P(z_{\mathrm{obs}_i}|\theta,u,\delta)\P(\theta,u,\delta)}\nonumber\\
&=& \frac{\P(\theta,u,\delta|z_i)\P(z_i)}{\P(u,\theta,\delta)}\frac{\P(z_{\mathrm{obs}_i}|\theta,u,\delta,z_i)}{\P(z_{\mathrm{obs}_i}|\theta,u,\delta)}\nonumber\\
&=&\P(z_i|\theta,u,\delta)\frac{\P(z_{\mathrm{obs}_i}|\theta,z_i,\delta,u)}{\P(z_{\mathrm{obs}_i}|u,\delta,\theta)}\nonumber\\
&=&\P(\theta,z_i|u)\frac{\P(\delta|\theta,z_i)}{P(\delta|u)}\frac{\P(z_{\mathrm{obs}_i}|\theta,z_i,\delta,u)}{\P(z_{\mathrm{obs}_i}|u, \delta,\theta)}
\end{eqnarray}
We assume $z_{\mathrm{obs}_i}$ is conditionally independent of $u$ given $z_i$ and $\delta$ and therefore we arrive at
\begin{equation}
\P(z_i|\theta, u, \delta, z_{\mathrm{obs}_i}) = \P(\theta,z_i|u)\frac{\P(\delta|\theta,z_i)}{\P(\delta|u)}\frac{\P(z_{\mathrm{obs}_i}|\theta,z_i,\delta)}{\P(z_{\mathrm{obs}_i}|\delta, \theta)}
\label{eq:zz_posterior}
\end{equation}
The first term on the right-hand side of the above equation describes the distribution of galaxies in redshift space. However, as we perform a real-space inference, we transform this term from redshift- $z$, to comoving Cartesian, $r$, space
\begin{equation}
\P(\theta,z_i|u) = \Big{|}r^2(z_i)\sin{(\phi)}\frac{\partial r}{\partial z}\Big{|}_{z_i}\Big{|}\P(\mathbf{x}|u),
\end{equation}
$\phi$ being the declination of the galaxy and $\mathbf{x}$ its real-space position. The real-space galaxy positions are used to build the gridded galaxy count field, $\M$. $\M$ indicates the galaxy counts except for the galaxy under consideration. We the last term on the right-hand side of the above equation as a marginal over galaxy counts
\begin{equation}
    \P(\mathbf{x}|u) = \sum_{\M} \P(\M|u)\P(\mathbf{x}|\M)
\end{equation}
The first term in the sum is the Nearest Grid Point projection that we use to grid galaxies at the field-level. We will indicate the Nearest Grid Point kernel with $W_\mathrm{NGP}$. Under the assumption that galaxies are Poisson-distributed, the number counts in each voxel are independent events. Since the Poisson intensity depends only on the density, our Poisson model is homogeneous. Following these assumptions
\begin{eqnarray}
    \P(\mathbf{x}|u) &=& \sum_{\M} \P(\mathbf{x}|\M)\nonumber\\ 
    &\times&\prod_i \delta^\mathrm{D}\left(\M_i-\sum_p W_\mathrm{NGP}(\mathbf{x}_i-\mathbf{x}_p)\right),
\end{eqnarray}
where $i$ is the voxel index and $p$ the galaxy index. We now want to rewrite the first term in the above sum with respect to the entire galaxy count field, $\N$. It differs from $\M$ in that it includes the galaxy under consideration. The reason we rewrite the redshift posterior with respect to $\N$ is because the entire galaxy count field is used in the density inference. Below this point we will drop the dependence on $u$, because all quantities are conditionally independent of $u$ given $\M$.
\begin{equation}
    \P(\mathbf{x}|\M) = \frac{1}{\P(\M)}\sum_{\N}\P(\N)\P(\mathbf{x}|\N)\P(\M|\N,\mathbf{x})
\end{equation}
Following our definition, the relationship between $\M$ and $\N$ is deterministic and yields
\begin{eqnarray}
    \P(\mathbf{x}|\M) &=& \frac{1}{\P(\M)}\sum_{\N}\P(\N)\P(\mathbf{x}|\N)\nonumber\\
    &\times& \prod_i \delta^\mathrm{D}\left(\M_i-[\N_i-W_\mathrm{NGP}(\mathbf{x}_i-\mathbf{x})]\right)\nonumber\\
    &=& \P(\mathbf{x}|\N),
\end{eqnarray}
because $\P(\N)$ and $\P(\M)$ are equal, as they differ by one (at the position of the galaxy under consideration). The above result is equal to finding a galaxy at position $\mathbf{x}$ given a number counts field. 
\begin{equation}
\P(\mathbf{x}|\N) = \sum_i \frac{W_\mathrm{NGP}(\mathbf{x}_i-\mathbf{x})}{\delta V}\frac{\N_i}{N_\mathrm{total}},
\label{eq:counts_position}
\end{equation}
where $\delta V$ is the volume of a voxel and $N_\mathrm{total}$ the total number of galaxies in the inference domain. As a result of the above, the second term on the right-hand side of \refeq{zz_posterior} is written as
\begin{equation}
    \frac{\P(\delta|\theta,z_i)}{\P(\delta|u)} = \prod_i \frac{M_i!}{N_i!}\lambda_i^{N_i-M_i} = \prod_i \left(\frac{\lambda_i}{N_i}\right)^{W_\mathrm{NGP}(\mathbf{x}_i-\mathbf{x})},
    \label{eq:density_positions}
\end{equation}
as it represents the ratio between the Poisson distribution for all galaxies and all galaxies expect for the one we consider at each step. Substituting \refeq{counts_position} and \refeq{density_positions} into \refeq{zz_posterior} we arrive at
\begin{eqnarray}
    \P(z_i|\theta, u, \delta, z_{\mathrm{obs}_i}) &\propto& \Big{|}r^2(z_i)\sin{(\phi)}\frac{\partial r}{\partial z}\Big{|}_{z_i}\Big{|}\nonumber\\
    &\times& \sum_i W_\mathrm{NGP}(\mathbf{x}_i-\mathbf{x})\lambda_i \P(z_{\mathrm{obs}_i}|\theta,z_i,\delta).
\end{eqnarray}
As demonstrated in \cite{Jasche:2013} the above form suggests that each galaxy can be sampled independently from all others, as the dependence on $u$ vanishes through the conditioning on galaxy counts and therefore, the Poisson intensity. Finally, since $\sin{(\phi)}$ is a constant term, it serves as a proportionality constant in each galaxy's target posterior and therefore vanishes. Finally, we drop the dependence on right ascension and declination for legibility, since we only sample redshifts. Therefore, the process in \refeq{z_sample} is equivalent to drawing a sample from
\begin{eqnarray}
    \P(z_i|\delta, z_{\mathrm{obs}_i}) &\propto& \Big{|}r^2(z_i)\frac{\partial r}{\partial z}\Big{|}_{z_i}\Big{|}\nonumber\\
    &\times& \sum_i W_\mathrm{NGP}(\mathbf{x}_i-\mathbf{x})\lambda_i \P(z_{\mathrm{obs}_i}|z_i,\delta).
\end{eqnarray}

\section{Estimators of large-scale structure properties}\label{app:data}

Accurate inferences of the large-scale structure with photometric galaxy clustering are necessary because next-generation photometric surveys probe deeper redshifts and have a wider footprint than spectroscopic surveys. Below, we discuss the aspects of the cosmic large-scale structure which we demonstrate our method on, as well as the estimators we use.

\subsection{Peculiar velocities}\label{pec_vel}

In order to derive the peculiar velocity field, we use the Simplex-In-Cell (SIC) estimator \citep[e.g.][]{2012MNRAS.427...61A,2015MNRAS.454.3920H, 2017JCAP...06..049L}. Contrary to kernel methods, which sample the velocity field only at a discrete set of locations with poor resolution in low-density regions, the SIC estimator provides an estimate of the velocity field at any point in space. In this framework, the Lagrangian positions of the particles in the inference domain are considered as vertices of unit cubes. The Delaunay tessellation of each unique cube defines six tetrahedra which are followed during the evolution. Subsequently, each tetrahedron deposits a value to the grid \citep[][Equation 37]{2017JCAP...06..049L}.

We further explore kinematic properties of the peculiar velocity field. Here, we focus on the peculiar velocity divergence, defined as
\begin{equation}
    \theta = -\frac{1}{fHa}\mathbf{\nabla} \cdot \mathbf{v},
\end{equation}
where $f$ is the linear growth rate, $H$ the Hubble parameter and $a$ the cosmic scale factor. The divergence determines how the volume of a peculiar velocity flow changes. The peculiar velocity divergence can be used to constrain $\Omega_m$ and structure formation \citep[e.g.][]{1995MNRAS.274...20B,2017MNRAS.471.3135H}. Further, the redshift-space distortion signal can be detected in the autocorrelation power spectrum of the peculiar velocity divergence and the cross-correlation power spectrum between the dark matter density and the peculiar velocity divergence \citep[e.g.][]{2019A&A...622A.109B}. For $\mathbf{\nabla} \cdot \mathbf{v}>0$ the region is expanding, for $\mathbf{\nabla} \cdot \mathbf{v}<0$ the region is contracting, whereas for $\mathbf{\nabla} \cdot \mathbf{v}=0$ the volume remains invariant. 

\subsection{Gravitational potential}\label{tidal_shear}

Probing gravitational tidal forces with photometric surveys is a challenging endeavor, but a necessary one \citep[e.g.][]{2015PhR...558....1T}, as intrinsic galaxy alignments contaminate weak lensing analyses and can be used as cosmological probes. Further, as photometric surveys extend deeper in redshift, they allow us to probe the evolution of intrinsic alignments over time. This evolution can be used to constrain galaxy formation and evolution scenarios.

For this purpose, we explore differing aspects of the dynamics and kinematics of the large-scale structure. We derive the gravitational potential of the dark matter density field by solving Poisson's equation in Fourier space
\begin{equation}
    \Phi(\mathbf{k}) = -\frac{4\pi G \rho(\mathbf{k})}{|\mathbf{k}|^2},
\end{equation}
where $G$ is the gravitational constant, $\rho$ the density and $\mathbf{k}$ the wavevector. We further derive the tidal shear of the gravitational potential, which is given by
\begin{equation}
T_{ij} = \frac{\partial^2\Phi}{\partial x_i \partial x_j},
\label{eq:tidal_shear}
\end{equation}
where $x_i$, ($\{i,j\} = \{1,2,3\}$) are comoving Cartesian coordinates. Our inference can further be used with cosmic web classification methods, particle- \citep[e.g.][]{2013arXiv1302.6221S} or cell-based \citep[e.g.][]{2018MNRAS.473.1195L,2020MNRAS.497.5041B} to infer cosmic structures constrained by photometric redshifts. Such classifications have been used to study the environment of cosmological tracers \citep[e.g.][]{2016JCAP...08..027L,2018A&A...612A..31P,2022MNRAS.510..366T}. \cite{2019A&A...625A.130K} further showed that photometric redshifts are positively correlated with filaments detected in spectroscopic galaxy observations.

\end{appendices}

%
\bibliographystyle{aa} 
\bibliography{aanda} 
%

\end{document}